\documentclass[preprint,12pt]{elsarticle}

\usepackage{amssymb}

\journal{Physics Letters A}

\begin{document}

\begin{frontmatter}



\title{On the Existence of Hydrogen Atoms in Higher Dimensional Euclidean Spaces}


\author[1,2]{F. Caruso}
\ead{francisco.caruso@gmail.com}
\author[2]{J. Martins} \author[2]{V. Oguri}
\address[1]{Centro Brasileiro de Pesquisas F\'{\i}sicas - Rua Dr. Xavier Sigaud, 150, 22290-180, Urca, Rio de Janeiro, RJ, Brazil}
\address[2]{Instituto de F\'{\i}sica Armando Dias Tavares, Universidade do Estado do Rio de Janeiro - Rua S\~ao Francisco Xavier, 524, 20550-900, Maracan\~a, Rio de Janeiro, RJ, Brazil}
\begin{abstract}

The question of whether hydrogen atoms can exist or not in spaces with a number of dimensions ($D$) greater than 3 is revisited. The lowest quantum mechanical stable states and the corresponding  wave functions are determined by applying Numerov's method to solve Schr\"odinger's equation. States for different angular momentum quantum number and dimensionality are considered. One is lead to the result that hydrogen atoms in higher dimensions could actually exist. The most probable distance between the electron and the nucleus are then computed as a function of $D$ showing the possibility of tiny confined states.

\end{abstract}


\end{frontmatter}


\section{Introduction}
\label{int}

Unification efforts based on superstring theories reinforced the idea that physical space dimensionality should be somehow explained in the framework of theoretical physics, even if some dimensions are supposed to curled up into a compact manifold~\cite{Barton}. A modern and comprehensive survey of dimensionality can be found in~\cite{Petkov}. Searches for experimental evidences of extra dimensions are part of contemporary tendencies for investigating physics ``beyond the Standard Model'' in Collider Physics~\cite{CMS}.

The idea that a particular physical law should depend on space dimensionality can be traced back to a philosophical speculation due to the young Kant~\cite{Kant}, namely, the suggestion that the Newtonian gravitational force depends on the three dimensionality of space. This insight was indeed proliferous and had inspired, for example, Ehrenfest, who was the first to give a mathematical ground to Kant's hint by discussing the mechanical stability of the formal solutions of the planetary motion in $D$ dimensions, assuming it is still described by a Poisson equation generalized to $R^D$, as well as the implications of higher dimensions on Bohr's atomic model~\cite{Ehrenfest17, Ehrenfest20}.

Those papers bring the belief that one can learn about space dimensionality from a class of generalized physical equations. In particular, this supposition was the
starting point of an article by Tangherlini~\cite{Tangherlini}, where the problem of Schr\"odinger's hydrogen atom in $D$ dimensions was formally treated for the
first time.
Although the basic assumption that one can \textit{prove} the three-fold nature of space from the generalized Schr\"odinger's equation
was shown to have epistemological limitations~\cite{Caruso}, the general ideas that it is possible to understand how the structure of a particular physical equation depends on dimensionality, or how dimensionality itself can depend on the spatial scale or even how a physical phenomenon could vary by changing the number of dimensions or other topological feature of space are still of considerable interest.

In a nutshell, the study of such a relationship between the structure of physical laws and space dimensionality is particularly attractive in the framework of the general theoretical scenario where extra space-time dimensions should play an important role on several attempts to unify the fundamental forces. Also the expectation from various scenarios of quantum gravity that the space-time dimension seems to rely on the size of the probed region, being somewhat smaller than four at small scales and monotonically raises with increasing the size of region \cite{Maziashvili} can be argued to justify the investigation on how physical laws depend on or are entangled in space-time dimensionality.
As a few examples of such a kind of dependence, one can quote the Casimir Effect \cite{Bender, Rafael, Neto}, the existence of scalar field lumps \cite{Gleiser} and how recent data on cosmic microwave background can be used to settle an upper limit for fractal space dimensions \cite{Oguri}. In any case, it could be possible to learn more about physics in 3 dimensions by investigating its generalization to spaces of higher dimensions.

\section{Former predictions}
\label{former_prediction}

 Are there stable hydrogen atoms in flat spaces if the number of dimensions is greater than three? Both positive and negative answers are found in the literature. One can try to summarize the main results as follows.

 In his semi-classical approach, Ehrenfest, based on Bohr quantization of circular atomic orbits, characterized by the total quantum number $n$, had shown  that for orbits in $D$ dimensions, the Bohr result for $D=3$,
$$E_n = - \frac{1}{2} \frac{m}{\hbar^2} \frac{e^4}{n^2}$$
should be replaced (for $D>2$) by
\begin{equation}\label{Ehrenfest_energy}
E_n^{(D)} = \frac{(D-4)}{2(D-2)} \times n^{-\frac{2D-4}{4-D}} \times \left(\frac{m}{\hbar^2}\right)^{\frac{D-2}{4-D}} \times e^\frac{4}{4-D}
\end{equation}

So, according to this result, the energy becomes positive for $D\geq 5$ and increases to infinity while the orbits draw closer and closer to the nucleus.

To the best of our knowledge, Tangherlini was the first, in 1963, to formally treat the problem of the hydrogen atom from the point of view of wave mechanics \cite{Tangherlini}. The problem was considered by him from the point of view of stability arguments and he did not solve the Schr\"odinger equation for an electrostatic potential energy of the type $V= e/(D-2) r^{D-2}$. The author limited himself to give a semi-classical argument based on the mechanical stability of the classical orbits. He immediately claimed that it is clear that for the cases $D>5$, \textit{the energy levels have a point of accumulation at minus infinity: i.e., $r = 0$  is not as regular point; and hence there are no stable bound states. The case $D = 4$, can also be excluded by standard arguments}. The same argument is sustained in Ref.~\cite{Andrew}. Nevertheless, to the best of our knowledge, this affirmative answer should be taken with reserve. Indeed, from the general theory of differential equations, it is well known that when in the equation
$$ y^{\prime\prime} + P(x) y^\prime + Q(x) y = 0$$
one or both coefficient functions $P(x)$ and $Q(x)$ fails to be analytic at $x_\circ$, this point is a singular point, and it is said to be \textit{irregular} so far $(x -~x_\circ) P(x)$ and $(x - x_\circ)^2 Q(x)$ are not analytic. Whenever this happens, one can safely conclude that there might not exist a Frobenius series solution for the differential equation \cite{Simmons}, and that is all. In fact, a counterexample can be found in \cite{Spruch, Malley} where analytic study of the dominant effects of a polarization potential that depends on $r$ (distance between an incident charge and a multi-electron atom) like $1/r^4$ was done. Corrections to the inverse sixth-power term were treated in \cite{Kleinman}. Indeed, it was shown that the Schr\"odinger equation for inverse fourth- and sixth-power potentials reduces to peculiar cases of the double-confluent Heun equation and its Ince's limit, respectively \cite{Figueiredo}.

In 1971, the same problem was investigated supporting the conclusion against the existence of stable atoms and atomic structures for $D>3$ with negative energy spectrum, although positive spectrum was not analyzed~\cite{Gurevich}.  In 2005, it was claimed that for one electron atoms in dimensions greater than three there is no normalized wave function corresponding to bound states \cite{Braga}.

However, Burgbacher and collaborators, in 1999, have shown that it is possible to have stable hydrogen atoms in higher dimensional non-compact spaces~\cite{Burgbacher}. They admitted that \textit{the kinetic energy in the Schr\"odinger operator has the usual form described by the $D$-dimensional Laplacian and that the electrostatic interaction in the Schr\"odinger equation has the same form irrespective of the spatial dimensions.} Instead, they are obliged to change the structure of Maxwell equations in higher dimensional spaces.  In anyway, so far Maxwell's theory is concerned, one should not forget Weyl's classical result which showed that only in a (3+1) dimensional space-times can such electromagnetic theory be derived from a simple gauge invariant action integral, having a Lagrangean density which is conformally invariant \cite{Weyl18, Weyl19, Weyl}.

 On the other hand, exact solutions for the radial Schr\"odinger equation in $D$ dimensions were found for potential of the type $V(r)= ar^{-1} + br^{-2} + c r^{-3} + dr^{-4}$ \cite{Khan}. In the paper \cite{Hall}, the discrete spectrum of the Schr\"odinger equation corresponding to the potential $V(r)\simeq 1/r$, is treated in $D$ dimensions. The authors considered also in details the Cornell potential, $V(r) = -a/r +br$. Algebraic solutions for the supersymmetric hydrogen atom with the potential $V(r)\simeq 1/r$ are given in  \cite{Kirchberg}.

Another sort of approach considers the possibility of confined hydrogen atoms in higher space dimensions by spherical cavities \cite{Navarro, Shaqqor}. The ground state energies were computed in impenetrable spherical cavities for different dimensions by using the Coulombian potential  $\sim 1/r$ \cite{Shaqqor}.

All those ideas have motivated several approaches to the problem but with conflicting conclusions. In many of them, the kinetic part of Schr\"odinger operator is generalized while the Coulombian energy potential is taken as in the case of three dimensions.

 On the other side, in Ref.~\cite{Tutik}, it was shown that the Dirac equation in $D$ dimensions, generalized for a Coulombian potential (which is a superposition of a Lorentz scalar term, $V(r) = -a/r$, and the temporal component of a Lorentz quadri-vector of the type $V(r) = -b/r$), has an exact solution. There are also a number of works that generalize Dirac equation for one electron atoms under the action of an attractive central potential. In many of them, the  Coulombian potential form in three dimensions, $V(r)\simeq 1/r$, is maintained no matter the value of $D$ \cite{Simsek}, \cite{Shi-Hai}, \cite{Jiang}. Those works can also be regarded as a further motivation to revisit the problem of the existence (or not) of hydrogen atoms in higher dimension spaces. Since in this Letter we are not considering external magnetic fields and since spin effects on the energy spectrum are known to be very small (of the order of $\alpha^2 \simeq 10^{-4}$) in the Pauli equation, this spectrum can be determined in a very good approximation by Schr\"{o}dinger equation.

Admitting the hydrogen atom to be described by a generalized Schr\"odinger equation, is it true that whenever one postulates the validity of Gauss law in $D$ dimension non-compact spaces -- which requires the replacement of the potential energy by a function that goes like $\sim 1/r^{D-2}$, in addition to the generalization of the kinetic part of Schr\"odinger's Hamiltonian operator --, the very existence of hydrogen atoms or their stability is lost? This is what will be investigated in the next Section.

\section{General results}\label{our_result}

In a flat (Euclidean) space, the radial Schr\"odinger equation in $D$ dimensions can be inferred from the $D$-dimensional Laplacian operator $\nabla^2_{(D)}$ in spherical coordinates, which is given by
$$\nabla^2_{(D)}= \frac{\partial^2}{\partial r^2} + \frac{D-1}{r} \frac{\partial}{\partial r} + \frac{1}{r^2} \Omega^2 $$
where $r$ is the radial coordinate, and $\Omega^2$ is the Laplacian operator on the unit hyper-sphere $R^{D-1}$.
The solution of the generalized Schr\"odinger equation, $\psi(\vec r)$, can be expressed in terms of $D-1$ angles as
$$ \psi(\vec r) = R(r) Y_\ell (\theta, \phi_1, \phi_2,\cdots \phi_{D-2})$$
The functions $Y_\ell$ are the Gegenbauer polynomials, which are the angular solutions of the equation
$$\Omega^2 Y_\ell= -\ell (\ell + D - 2) Y_\ell,$$
while the radial solution $R(r)$ satisfies
\begin{eqnarray}
\label{radial_equation}
\nonumber 
  & & \frac{\mbox{d}^2R(r)}{\mbox{d}r^2} + \frac{D-1}{r} \frac{\mbox{d} R}{\mbox{d} r} + \\
  \, &+& \frac{2m}{\hbar^2} \left[E - U_{(D)}(r) - \frac{h^2}{2m} \frac{\ell (\ell + D - 2)}{r^2} \right] R = 0
\end{eqnarray}

\noindent where $U_{(D)} = - e_{(D)} V_{(D)}(r)$.

Making the choice analogous to the three dimensional case,
\begin{equation}
\label{R_definition}
u(r) = r^{(D-1)/2} R(r)
\end{equation}
the first order derivative can be eliminated, giving rise to
\begin{equation}
\label{D_radial_eq}
\frac{\mbox{d}^2u(r)}{\mbox{d}r^2} + \frac{2m}{\hbar^2} \left[\left( E - U_{\mbox{\tiny eff}} (r) \right) \right] u(r) = 0
\end{equation}

\noindent where the effective potential energy is given by the electrostatic energy plus the generalized centrifugal term
\begin{equation}
\label{effective_potential}
U_{\mbox{\tiny eff}} (r) = U_{(D)} +  \frac{\hbar^2}{2m} \frac{j(j+1)}{r^2}
\end{equation}
 with $m$ and $e_{(D)}$ being, respectively, the mass and the charge of the electron and the values of $j$ depend on the orbital angular momentum quantum number $l$ and on the number of dimensions $D$ as
\begin{equation}
\label{j_def}
j = l + \frac{D-3}{2}
\end{equation}

Up to this point there is a consensus, but the choice of $V(r)$ is not unique in the literature. Often, the usual Coulombian potential established for $D=~3$ is assumed to be valid for an arbitrary $D$. Instead, we will use throughout this Letter the formula that guarantees the electric charge conservation in a $D$ dimensional space, or, equivalently, the generalized Poisson equation, where the power of the Laplacian operator is maintained and just the number of independent spatial coordinates varies, \textit{i.e.},
\begin{equation}
\label{generalized_potential}
V_{(D)}(r) = \frac{2 \Gamma(D/2)}{\pi^{(D-2)/2}}\, \frac{e_{(D)}}{(D-2)r^{D-2}}
\end{equation}

Once the potential is fixed, it is straightforward to see that it does not belong to the class of potential to which the Schr\"odinger equation is exactly solved \cite{Lemieux}. Therefore, at this point, Eq.~(\ref{D_radial_eq}) will be solved for the potential energy given by (\ref{effective_potential}) and (\ref{generalized_potential}) by applying the Numerov numerical method \cite{Numerov_1, Numerov_2, Leroy} to solve an eigenvalue differential equation.
A specific program was developed by the authors in $C++$ language and both calculations and graphics were done by using the CERN/ROOT package.

In this Letter, the Schr\"odinger equation is solved numerically for few values of angular momentum, i.e., $\ell = 0,1,2,3$, allowing space dimensionality to vary between the interval $5 \leq D \leq 10$.

Before presenting the numerical results, one might say that, so far concerning the problem of space dimensionality, one should adopt something similar to the cosmological principle, according to which one should expect that the laws of physics, as determined in our neighborhood, are valid in all regions of the Universe and in all moments of its history, independently of the space-time scale which is being probed. An analogous hypothesis should be assumed here as in all literature, namely, that the physical law established in $D=3$ will be valid for different values of the dimensionality $D$ and that the numerical values of the physical constants should not vary significantly. Otherwise it will be impossible to get any numerical prediction.

The present calculations depend on the numerical value of the electric charge, $e_{(D)}$, in $D$ dimensions. It will be assumed throughout this Letter that $e_{(D)}$ has the same value of $e$ measured in three dimensions. This should be justified from the results of ref.~\cite{Nasseri}, where it is shown that the numerical value of the generalized fine structure function for $D$ dimensions is very close to the three dimensional one, $1/137$, and also Planck constant $\hbar_{(D)}$ did not vary significantly with space dimensionality.

\begin{figure}[htb!]
\includegraphics[width=100mm]{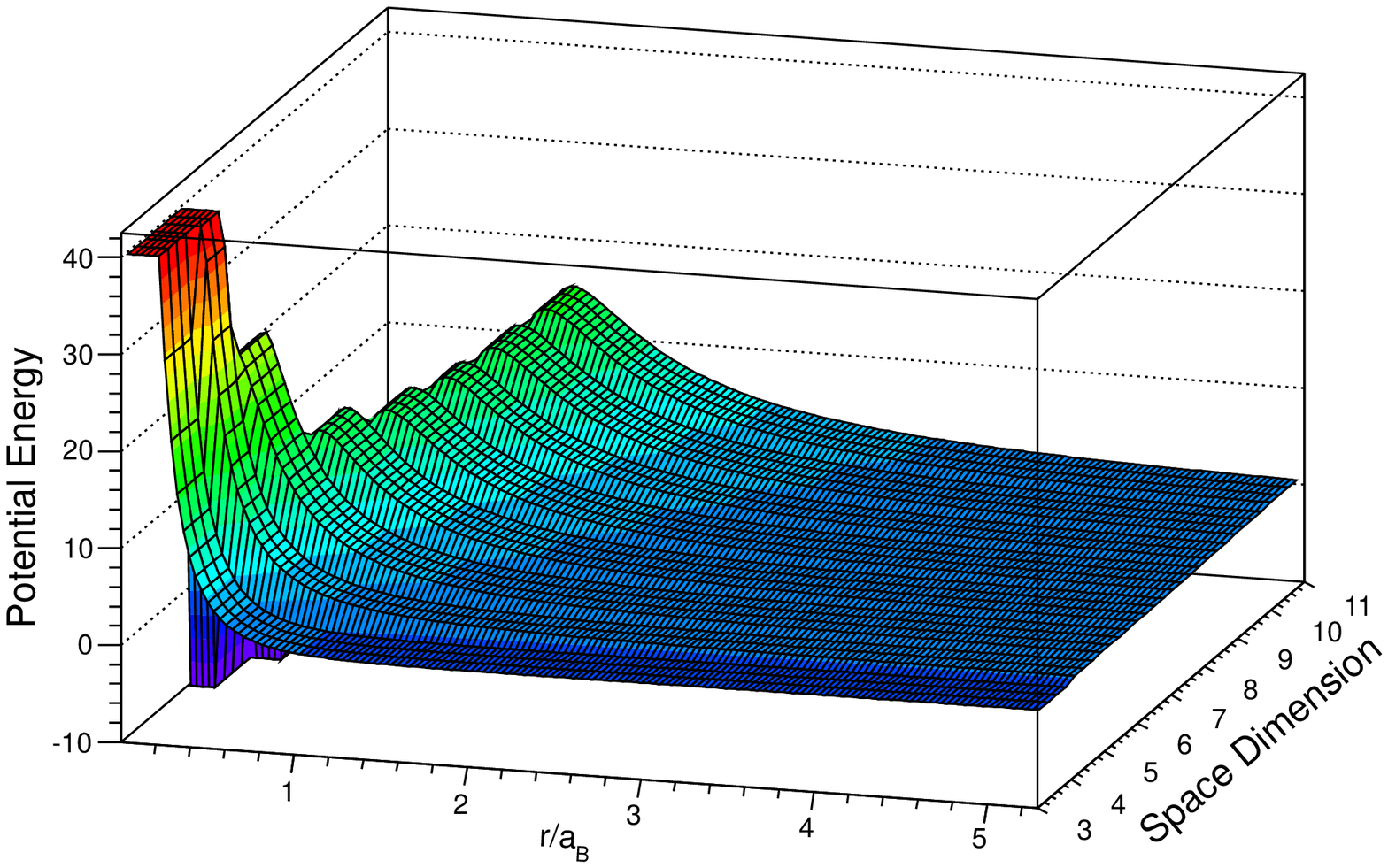}
\caption{Potential energy of the $\ell =1$ state of a hydrogen atom as a function of $r/a_B$ for the space dimensionality $3\leq D \leq 10$.}
\label{fig.1}
\end{figure}

With this assumption, the effective potential energy is given by fig.~\ref{fig.1} and the energy eigenvalues are calculated for the first four principal quantum numbers ($n=1,2,3,4$) and the lowest orbital angular momentum $\ell = 0,1$, varying $D$ from 5 to 10. The results are shown in table~\ref{tab.1} and table~\ref{tab.2}.

\begin{table}[ht]
\caption{Energy eigenvalues corresponding to principal quantum numbers $n=1,2,3,4$ and $\ell = 0$, for dimensions $5 \leq D \leq 10$.}
\label{tab.1}
\begin{center}
{\small
\begin{tabular}{c|c|c|c|c|c|c}
\hline\hline
$E_n$ [eV]     & $D$ = 5 & 6 & 7 & 8 & 9 & 10  \\
\hline
$E_1$       & 0.31  & 0.41    & 0.51     & 0.62   & 0.74  & 0.88  \\
$E_2$       & 1.77   & 2.05     & 2.30   & 2.57   & 2.86  & 3.14  \\
$E_3$       & 4.39   & 4.86   & 5.30   & 5.72    & 6.12  & 6.56   \\
$E_4$       & 8.16    & 8.88   & 9.49   & 10.03   & 10.63 & 11.20 \\
\hline\hline
\end{tabular}
}
\end{center}
\end{table}

\begin{figure}[htb!]
\includegraphics[width=120mm]{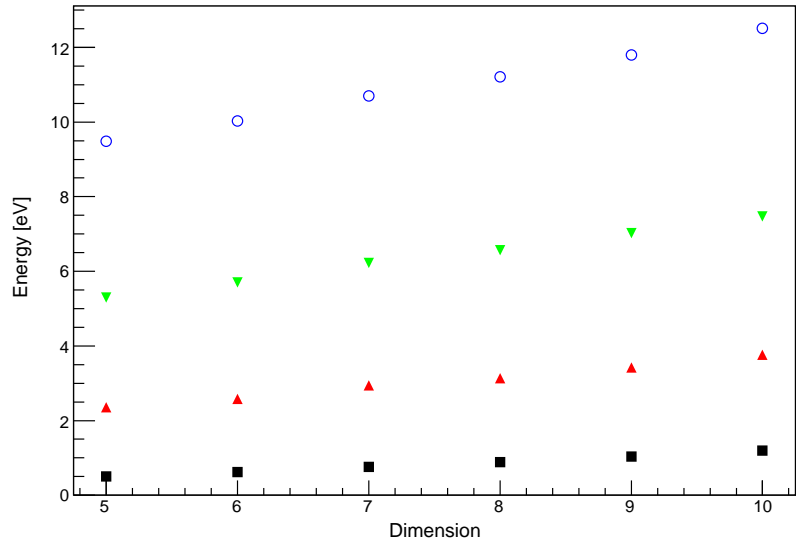}
\caption{Energy eigenvalues for $\ell =1$ state of an hydrogen atom as a function of space dimensionality. The black squares refer to the $E_1$ eigenvalues; the red triangles to the $E_2$; the green, to $E_3$ and the blue open circles to $E_4$.}
\label{fig.2}
\end{figure}

\begin{table}[ht]
\caption{Energy eigenvalues corresponding to principal quantum numbers $n=1,2,3,4$ and $\ell = 1$, for dimensions $5 \leq D \leq 10$.}
\label{tab.2}
\begin{center}
{\small
\begin{tabular}{c|c|c|c|c|c|c}
\hline\hline
$E_n$ [eV] & $D$ = 5 & 6 & 7 & 8 & 9 & 10  \\
\hline
$E_1$     & 0.51  & 0.61  & 0.76  & 0.89  & 1.04    & 1.20   \\
$E_2$     & 2.28  & 2.58  & 2.85  & 3.14  & 3.42    & 3.77  \\
$E_3$     & 5.25  & 5.70  & 6.13  & 6.57  & 7.03    & 7.47  \\
$E_4$     & 9.40  & 10.06 & 10.63 & 11.21 & 11.80   & 12.51 \\
\hline\hline
\end{tabular}
}
\end{center}
\end{table}
How these lowest energy eigenvalues (corresponding to the states $\ell =1$) grow with the increasing of the space dimensionality is shown in fig.~\ref{fig.2}.

It is important to stress that all those values are very small compared to the maximum value of the effective potential ($U_m$) for each value of $D$. Indeed, considering the $\ell =1$ state, the ratio $E_1/U_m$ goes from 0.1\% to 0.6\% when $D$ goes from~5 to 10. These results suggest a very low probability of tunneling effect.

In addition, we studied how the energy spectrum depends on $\ell$, for a particular $D$. For example, for $D=6$ and $\ell=0,1,2,3$, the energy eigenvalues are shown in table~\ref{tab.3}.

\begin{table}[ht]
\caption{Energy eigenvalues corresponding to principal quantum numbers $n=1,2,3,4$ and $\ell = 0,1,2,3$, for $D =6$.}
\label{tab.3}
\begin{center}
{\small
\begin{tabular}{c|c|c|c|c}
\hline\hline
$E_n$ [eV]     & $L$ = 0 & $L$ = 1 & $L$ = 2 & $L$ = 3  \\
\hline
$E_1$       & 0.410   & 0.615    & 0.976    & 1.085    \\
$E_2$       & 2.051   & 2.582    & 3.183    & 3.406    \\
$E_3$       & 4.864   & 5.702    & 6.664    & 6.971    \\
$E_4$       & 8.878   & 10.057    & 11.235    & 11.666    \\
\hline\hline
\end{tabular}
}
\end{center}
\end{table}

\begin{figure}[htb!]
\includegraphics[width=120mm]{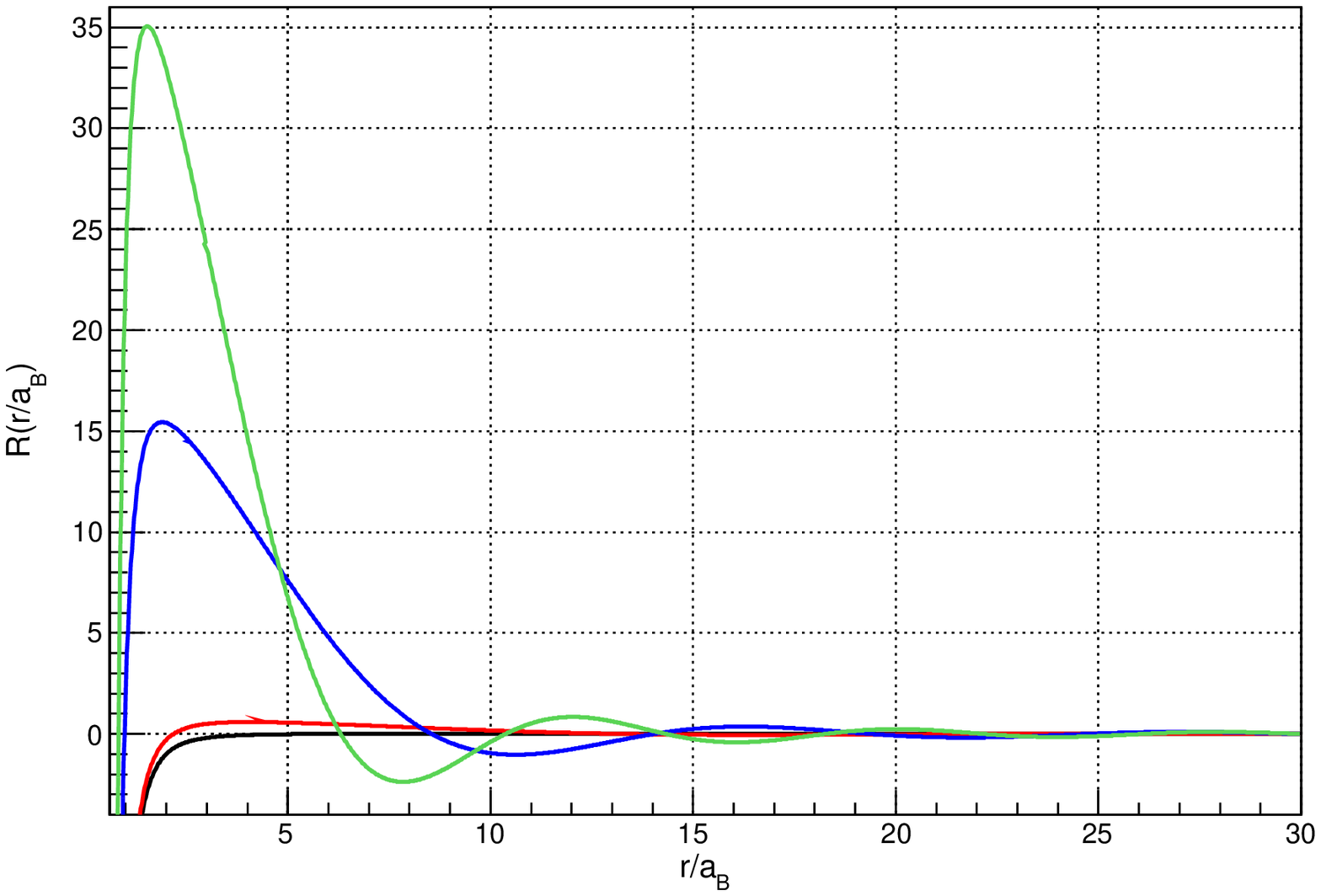}
\caption{Radial wave function $R_{\ell =0}$ for the first four states of the hydrogen atom in the case $D=6$. The color legend is: black curve refers to $E_1$ eigenvalue; red to $E_2$; blue to $E_3$ and green to $E_4$.}
\label{fig.3}
\end{figure}

\begin{figure}[htb!]
\includegraphics[width=120mm]{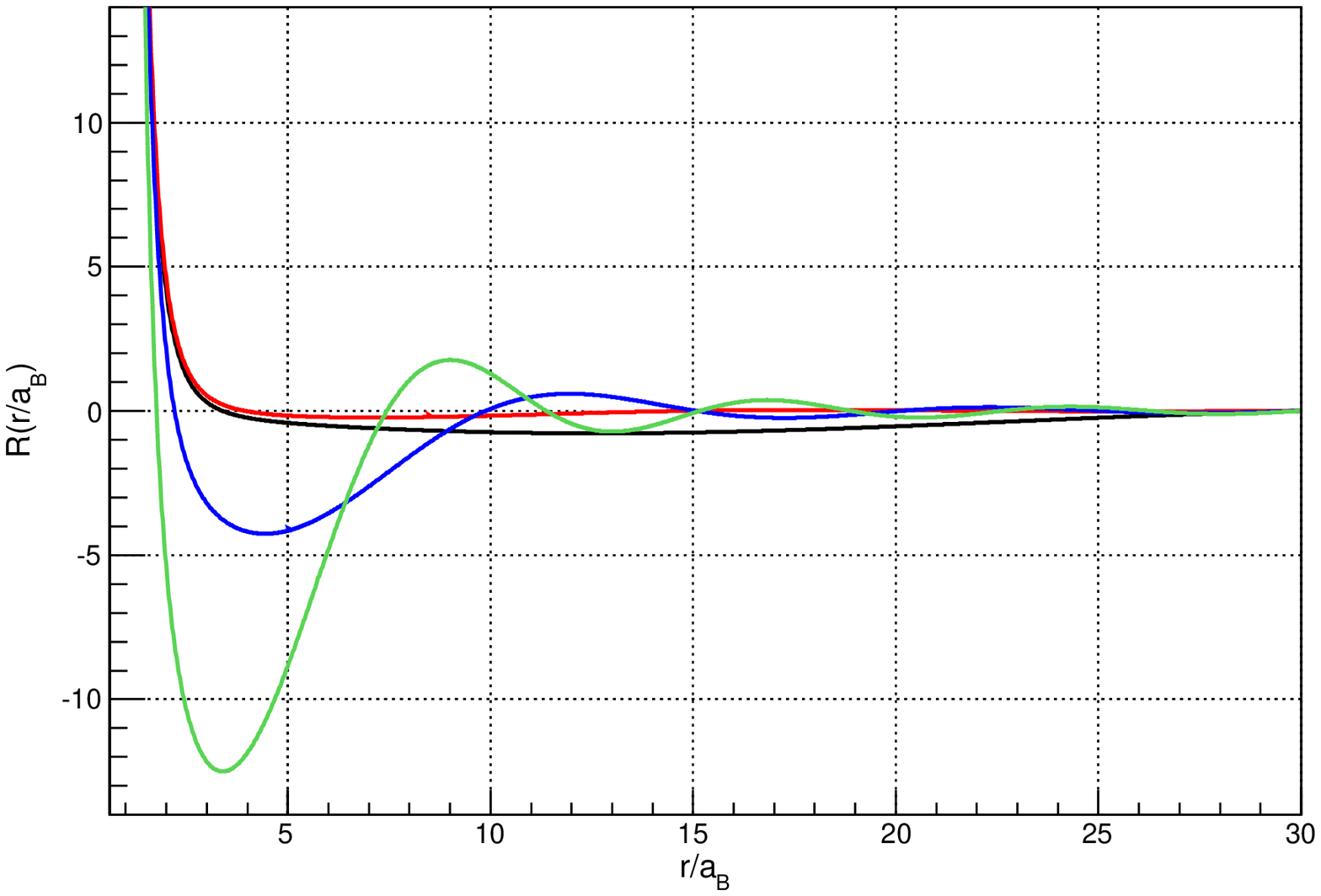}
\caption{Radial wave function $R_{\ell =1}$ for the first four states of the hydrogen atom in the case $D=6$. The color legend is: black curve refers to $E_1$ eigenvalue; red to $E_2$; green to $E_3$ and blue to $E_4$.}
\label{fig.4}
\end{figure}

\begin{figure}[htb!]
\includegraphics[width=120mm]{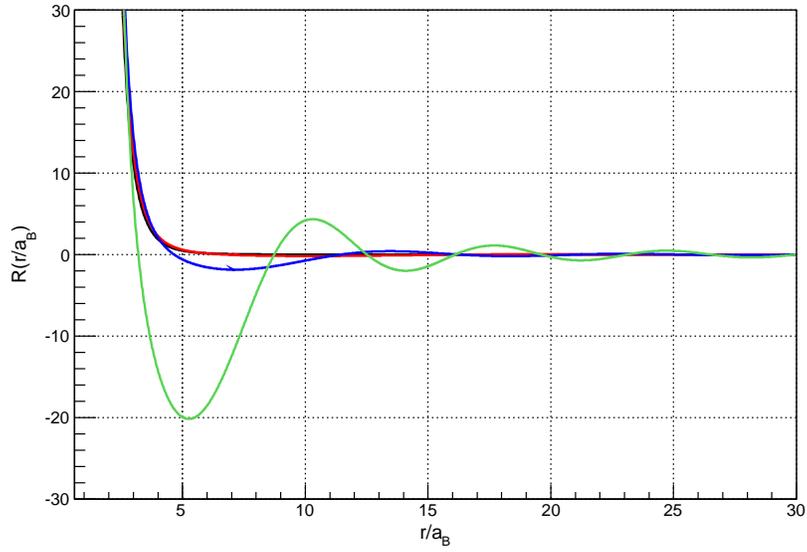}
\caption{Radial wave function $R_{\ell =3}$ for the first four states of the hydrogen atom in the case $D=6$.}
\label{fig.5}
\end{figure}

All the regular radial wave functions were determined and just few of them are plotted (with arbitrary normalizations) in Fig.~\ref{fig.3} ($\ell=0$), Fig.~\ref{fig.4} ($\ell=1$) and Fig.~\ref{fig.5} ($\ell=3$), all of them for the particular value $D=6$.

Once the radial wave functions, the mean value of the radius of the electron $\langle r \rangle$ for a particular dimensionality $D$, given by

\begin{equation}
\label{mean_r}
\langle r \rangle = \int r R_{\ell}^2 (r) r^{D-1}\, \mbox{d} r\, \mbox{d} \Omega_{D-1}
\end{equation}
was numerically calculated by solving the above integral by using the rejection Monte Carlo Method.

Just the predicted values for $\langle r \rangle$ and the respective errors (for the state $\ell =1$) as a function of $D$ are shown in table~\ref{tab.4}, where the values are expressed in terms of the numerical value of the Bohr radius $a_B$, equal to $0.529 \times 10^{-8}$~cm. It should be noticed that for the quantity $\langle r \rangle/a_B$ is multiplied by a factor 5 going from $D=5$ to $D=10$.


\begin{table}[h!]

\caption{Dependence of $\langle r \rangle$ upon the dimension $D$, for $\ell=1$ states. Errors for each prediction are shown in the last column.}
\label{tab.4}
\begin{center}
\begin{tabular}{c|c|c}

\hline\hline
$D$     & $\langle r \rangle$/a$_{B}$   & $\sigma_{r/a_{B}}$ \\

\hline
5       & 0.045     & 0.001 \\
6       & 0.090     & 0.002 \\
7       & 0.120     & 0.002 \\
8       & 0.174     & 0.003 \\
9       & 0.223     & 0.003 \\
10      & 0.252     & 0.003 \\
\hline\hline
\end{tabular}
\end{center}
\end{table}

\begin{figure}[htb!]
\includegraphics[width=120mm]{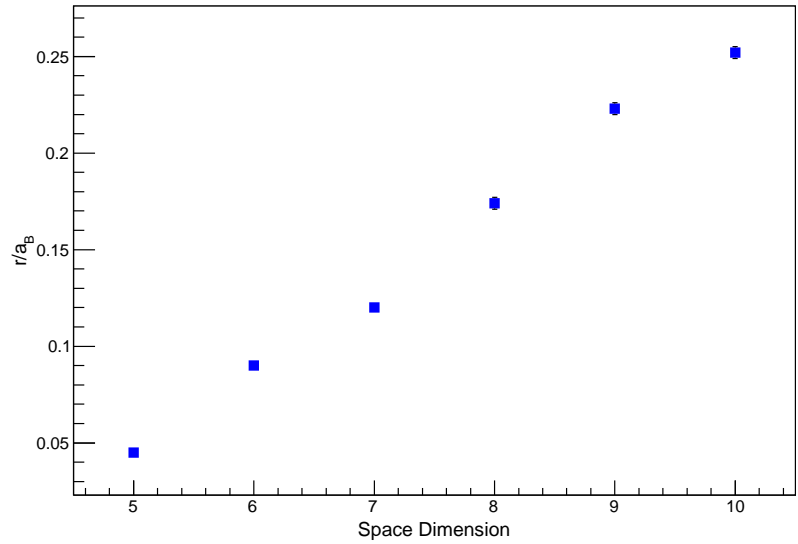}
\caption{Dependence of the mean values of the ground state ($\ell=1$ states) hydrogen atom radius upon space dimensionality.}
\label{fig.6}
\end{figure}

In fig.~\ref{fig.6} it is plotted the dependence of $ \langle r \rangle$ (normalized to the Bohr radius in $D=3$) on $D$ in the case $\ell = 1$.

The mean value of the distance between the electron and the proton in the hydrogen atom tends to grow by increasing the number of dimensions for $D \ge 5$ (for $D=4$ there is no state). This means that the distance between the electron and the proton, for $D=5$, is an order of magnitude greater than in the case $D=3$. Notice that this result is qualitatively different to Ehrenfest's statement that the orbits draw closer and closer to the nucleus as the value of $D$ increases.

\section{Final comments}\label{disc}

In this Letter it was shown that there is no state of a hydrogen atom with negative eigenvalues when space dimensionality is higher than 3. It is important to point out that for $D=4$ there are no solutions for this kind of atom. Additionally, for dimensions $D \ge 5$, it is argued that there are stable states with positive energy, described by well behaved wave functions, corresponding to an electron confined in a potential created by the central proton. The first energy levels are shown to be very small compared to the barrier high in order the electron could be confined in the proton potential, and the mean distance between the electron and the nucleus is about ten times greater than the correspondent distance in three dimensions.

The amount of energy necessary to ionize the atom in higher dimensions may be numerically estimated by our method, by computing all possible stable states with energies less than the maximum value of the confined potential generated by the central proton. For instance, see table~\ref{tab.1}, for $D=5$, the value of $E_4$ is indeed the biggest one still less than the maximum value of the potential. Thus, in this case, one can estimate the ionization energy as something close to this value. For other values of space dimensionality, more states fulfilling this condition can be found, even if they were not shown in table~\ref{tab.1}.

In summary, one can ask what could be learned from the results presented here. Admitting the hydrogen atom is described by the generalized Schr\"{o}dinger equation no matter what space dimensionality is, it was found that now there are stable mathematical solutions not only for $D=3$, against to what was claimed by different authors in the past. Indeed, it was shown that there are stable states even for higher dimensional Euclidean spaces, but all with positive energies. Therefore, the experimental observation that energy of the bound state hydrogen atom is of the order of $-13.6$~eV seems to indicate that nature should somehow prefer tridimensional space.





\bibliographystyle{elsarticle-num}



\newpage

\newpage

\newpage

\newpage

\newpage

\newpage

\end{document}